\documentclass[aps,pre,twocolumn,10pt,amsmath,amssymb,amsfonts,superscriptaddress]{revtex4-1}
\usepackage                   {graphicx}
\usepackage[usenames]         {color}
\usepackage[colorlinks=True,
            urlcolor=myblue,
            linkcolor=myblue,
            citecolor=mygreen]{hyperref}

\definecolor{mygreen}{RGB}{ 0, 127,   0}
\definecolor{myblue} {RGB}{ 0,   0, 127}
\newlength{\figwidth}
\newcommand{\fig}[1]{Fig.~\ref{#1}}
\newcommand{\eq}[1]{Eq.~(\ref{#1})}
\newcommand{\ud}{\,\mathrm{d}}
\newcommand{\vect}[1]{\mathbf{#1}}

\begin{document}
\title{A pair potential that reproduces the shape of isochrones in molecular liquids}
\author{Arno A. Veldhorst}\email{a.a.veldhorst@gmail.com}
\affiliation{``Glass and Time'', IMFUFA, Department of Science and Environment, Roskilde University, P.O. Box 260, DK-4000 Roskilde, Denmark}
\affiliation{Laborat\'orio de Espectroscopia Molecular, Instituto de Qu\'imica, Universidade de S\~ao Paulo, CP 26077, CEP 05513-970 S\~ao Paulo, SP, Brazil}
\author{Thomas B. Schr{\o}der}\email{tbs@ruc.dk}
\affiliation{``Glass and Time'', IMFUFA, Department of Science and Environment, Roskilde University, P.O. Box 260, DK-4000 Roskilde, Denmark}
\author{Jeppe C. Dyre}\email{dyre@ruc.dk}
\affiliation{``Glass and Time'', IMFUFA, Department of Science and Environment, Roskilde University, P.O. Box 260, DK-4000 Roskilde, Denmark}
\date{\today}

\begin{abstract}
Many liquids have curves (isomorphs) in their phase diagrams along which structure, dynamics, and some thermodynamic quantities are invariant in reduced units. A substantial part of their phase diagrams is thus effectively one dimensional.
The shape of these isomorphs is described by a material-dependent function of density $h(\rho)$, which for real liquids is well approximated by a power law $\rho^\gamma$.
However, in simulations, a power law is not adequate when density changes are large; typical models such as the Lennard-Jones liquid show that $\gamma(\rho) \equiv \ud \ln h(\rho)/\ud \ln \rho$ is a decreasing function of density.
This paper presents results from computer simulations using a new pair potential that diverges at a nonzero distance and can be tuned to give a more realistic shape of $\gamma(\rho)$. 
Our results indicate that the finite size of molecules is an important factor to take into account when modeling liquids over a large density range.
\end{abstract}

\maketitle

It is still an open question what controls the dynamics of viscous, glass-forming liquids~\cite{Debenedetti2001, Dyre2006, Lubchenko2007, Kivelson2008, Cavagna2009, Ediger2012}. Although the dynamics in general depends on both temperature $T$ and density $\rho$, the dynamics of many organic supercooled liquids can be collapsed onto a single curve when plotted against a combined, material-specific variable $h(\rho)/T$~\cite{Tolle1998, Alba-Simionesco2002, Tarjus2004}. It was found in many experiments that the scaling function $h(\rho)$ is generally well approximated by a power law as $h(\rho)=\rho^\gamma$, with $\gamma$ being the material-specific density-scaling exponent~\cite{Roland2005, Roland2010}; we refer to this as \emph{power-law density scaling}. Another important development was the discovery that the dynamics of liquids are a function of the excess entropy~\cite{Rosenfeld1977,Dzugutov1996}.

The isomorph theory~\cite{paper4} explains why both density scaling and excess-entropy scaling work for some liquids. Liquids that obey the isomorph theory have curves in their phase diagram, so-called isomorphs, along which not only the dynamics, but also the structure, excess entropy and other thermodynamic quantities are invariant. The development of the isomorph theory was initiated by the observation that in computer simulations some liquids have strongly correlated fluctuations in their energy and pressure. More specifically, if the energy $E$ and pressure $p$ are split in a kinetic part and a configurational part that only depends on the particle positions $\vect{R}\equiv(\vect{r}_1,\ldots,\vect{r}_N)$, as follows
 \begin{equation}\begin{aligned}
  E  &= K + U(\vect{R})\,,\\
  pV &= N k_B T + W(\vect{R})\,,
\end{aligned}\end{equation}
the strong correlations are found between the thermal equilibrium fluctuations of the potential energy $U$ and the virial $W$ in the $NVT$ ensemble~\cite{Pedersen2008}, although strong correlations have also been found at high pressures in the $NpT$ ensemble~\cite{Coslovich2009}. Indeed, the standard correlation coefficient 
\begin{equation}\label{eq:R}
  R = \frac{\left\langle \Delta W \Delta U \right\rangle}
           {\sqrt{\left\langle(\Delta W)^2\right\rangle \left\langle(\Delta U)^2\right\rangle}}
\end{equation}
indicates whether a liquid obeys the isomorph theory: this is the case whenever $R>0.9$ (although this value is of course somewhat arbitrary). The standard linear regression ``slope'' of the fluctuations
\begin{equation}\label{eq:gamma-fluctuations}
  \gamma \equiv \frac {\left<\Delta W \Delta U\right>} {\left<(\Delta U)^2\right>}
\end{equation}
is the density-scaling exponent~\cite{paper4}, and the theory thus provides a convenient way to determine the density-scaling exponent in computer simulations.

Another empirical observation that can be explained by the theory is that many liquids have been shown to obey isochronal superposition, meaning that the relaxation spectra of a liquid have identical shapes if the average relaxation time is the same~\cite{Tolle2001, Roland2003, Ngai2005}. Also several phenomenological melting rules can be explained by the isomorph theory~\cite{Gilvarry1956, Ross1969, Saija2006}. The Lindemann melting criterion for instance states that crystals melt when the atomic vibrational displacement reaches a certain value in reduced units. This follows from the theory since the melting line is predicted to be an isomorph to a good approximation~\cite{paper4}, and indeed many other properties have also been found to be isomorph invariants along the melting line~\cite{Heyes2015, Heyes2015a, Costigliola2016a}.

Many model systems studied so far have been shown to obey the isomorph theory, including atomic liquids with a range of different pair potentials~\cite{paper1, Veldhorst2012, Bacher2014a, Veldhorst2015}, crystals~\cite{Albrechtsen2014}, as well as rigid~\cite{Ingebrigtsen2012b} and flexible molecular liquids~\cite{Veldhorst2014}. Experimental evidence for the isomorph theory proved harder to get, but has been provided as well~\cite{paper2, Gundermann2011, Xiao2015}. For a detailed description of the isomorph theory focusing on its validation in simulations and experiments, the reader is referred to a recent Feature article~\cite{Dyre2014}.

The isomorph theory has been tested most thoroughly in computer simulations, making it possible to investigate very large ranges of density.
Interestingly, most models that have been simulated show that power-law density scaling does not work, i.e., the scaling function $h(\rho)$ which describes the shape of the isomorphs, is \emph{not} a power law~\cite{paper1, Veldhorst2012, Ingebrigtsen2012b, Veldhorst2014, Bacher2014a, Albrechtsen2014, Veldhorst2015}. 
Instead, the scaling function $h(\rho)$,  which describes the isomorphs via the equation $h(\rho)/T={\rm Const.}$, is a more general function of density, and depends on the pair potential.
Only inverse-power-law (IPL) potentials with $\upsilon(r) \propto r^{-n}$ obey power-law density scaling exactly (with $\gamma = n/3$) and have perfectly correlated fluctuations in $U$ and $W$ ($R=1$).
For other model liquids, $h(\rho)$ is not known analytically although it can be determined from a single simulation if the pair potential is a sum of IPLs, as is the case for the well-known Lennard-Jones (LJ) pair potential~\cite{Ingebrigtsen2012, Bohling2012}.
In that case each power-law term $n$ in the potential leads to a term in $h(\rho)$, with the relative contribution of that term to the excess heat capacity $C^{ex}_V=\sum_n C^{ex}_{V,n}$ as the prefactor~\cite{Bohling2012}
\begin{equation}
  h(\rho) = \sum_n \dfrac{C^{ex}_{V,n}}{C^{ex}_V} \rho^{n/3}\,.
\end{equation}
The relative contributions of each term in the potential to $C^{ex}_V$ can easily be determined from a single simulation.

Thus, there is a discrepancy between the simulations that show that power-law density scaling does not work for most model systems and experiments that show it gives satisfactory collapse for most nonassociating liquids. It is not clear what causes this discrepancy. On the one hand, one could argue that the models used in simulations are too simple to properly capture the physics of real liquids. However, this does not explain why IPL potentials, arguably the simplest pair potentials, \emph{do} predict the power-law density scaling seen in so many different liquids. Indeed, there is no a priori reason why $h(\rho)$ should be a power law, it is used mostly for empirical reasons.

 On the other hand, the discrepancy could be explained by the fact that it is much easier to obtain a large range of densities in simulations than it is in experiments. This hypothesis has led to more experimental investigations into liquid dynamics over larger ranges of density~\cite{Bohling2012, Abramson2014, Casalini2016a, Casalini2016b}. B{\o}hling \emph{et al.}~\cite{Bohling2012} found that dibutylphthalate (DBP) and decahydroisoquinoline (DHIQ) do not conform to power-law density scaling, although the relaxation time data could still be collapsed when another $h(\rho)$ was used. However, the validity of this analysis is still under discussion as old and possibly outdated equations of state were used, leading to large extrapolations to determine densities. More recent studies claim to get a good collapse of the data with power-law density scaling for both DHIQ~\cite{Casalini2016a} and DBP~\cite{Casalini2016b}.

Simulation data and the analysis by B{\o}hling \emph{et al.} find an $h(\rho)$ that is not a power law, meaning that the logarithmic slope
\begin{equation}\label{eq:gamma-hrho}
  \gamma(\rho) = \dfrac{\ud \ln h(\rho)}{\ud \ln \rho},
\end{equation}
which gives the ``local'' density scaling exponent, is not constant. Computer simulations of various pair potentials~\cite{paper1, Veldhorst2012, Bacher2014a} and molecular systems~\cite{Schroder2009a, Ingebrigtsen2012b, Veldhorst2014} have found $\gamma(\rho)$ to be a decreasing function of density. The fact that the LJ potential has a decreasing $\gamma(\rho)$ can be understood by considering that at high densities, particles are close to each other and only feel the repulsive $r^{-12}$ term. For high densities, $\gamma$ should thus approach $12/3=4$. At normal densities (around zero pressure), the attractive $r^{-6}$ term plays an important role, however, and because it is subtracted it makes the Lennard-Jones potential steeper than the $r^{-12}$ IPL. Therefore, at low pressure, the LJ potential has a higher scaling exponent than expected from its $r^{-12}$ term, and one finds here that $\gamma \approx 6$~\cite{paper1}.

Although there is no consensus on whether $h(\rho)$ is always well described by a power law~\cite{Bohling2012, Casalini2016a, Casalini2016b}, none of the experimental data are in agreement with the fact that $\gamma(\rho)$ decreases with density for standard pair potentials such as the LJ~\cite{paper1} and Buckingham~\cite{Veldhorst2012} potentials. This led us to investigate other pair potentials for which the density scaling exponent does not decrease with increasing density.

As the addition of the attractive term makes the $\gamma(\rho)$ a decreasing function in the Lennard-Jones potential, we instead start with a potential that has an increasing $\gamma(\rho)$. We note that such a potential would mean that particles would feel a steeper repulsion as the interparticle distance decreases. This led us to suggest a sum of infinitely many power laws;
\begin{equation}\label{integral}
  \int\limits_p^\infty r^{-n}\mathrm{d}n
\end{equation}
as a candidate, as the higher exponents would dominate at shorter $r$. For it to be a more realistic model of a molecular liquid, we also include an attraction as
\begin{equation}
  A\int\limits_p^\infty r^{-n} \mathrm{d}n - B\int\limits_q^\infty r^{-n} \mathrm{d}n
  = A\dfrac{r^{-p}}{\ln(r)} - B\dfrac{r^{-q}}{\ln(r)}\,,
\end{equation}
with $p>q$. This potential diverges at $r_0=1$, but to easily set the potential minimum $r_m$, the divergence diameter $r_0$, and the potential depth $\varepsilon$ we parameterize the potential as
\begin{equation}\label{eq:alog_potential}
  \upsilon(r) = 
  \varepsilon\: \dfrac{A\left(r/r_m\right)^{-p} - B\left(r/r_m\right)^{-q}}
                      {\ln\left(r/r_0\right)}\,,
\end{equation}
where
\begin{align*}
  A &= \dfrac{q \ln\left(r_m/r_0\right) + 1}{p-q}\,,\\
  B &= \dfrac{p \ln\left(r_m/r_0\right) + 1}{p-q}\,.
\end{align*}
In this study, we choose the exponents $p=12$ and $q=6$, like the standard LJ (LJ12-6) potential, and we designate it as the S12-6 potential due to the integral in \eq{integral}. The potential reduces to the standard LJ potential in the limit of $r \rightarrow 0$. In the limit of $r_0 \rightarrow r_m$ the potential becomes purely attractive.

\begin{figure}
  \centering
  \includegraphics[width=\figwidth]{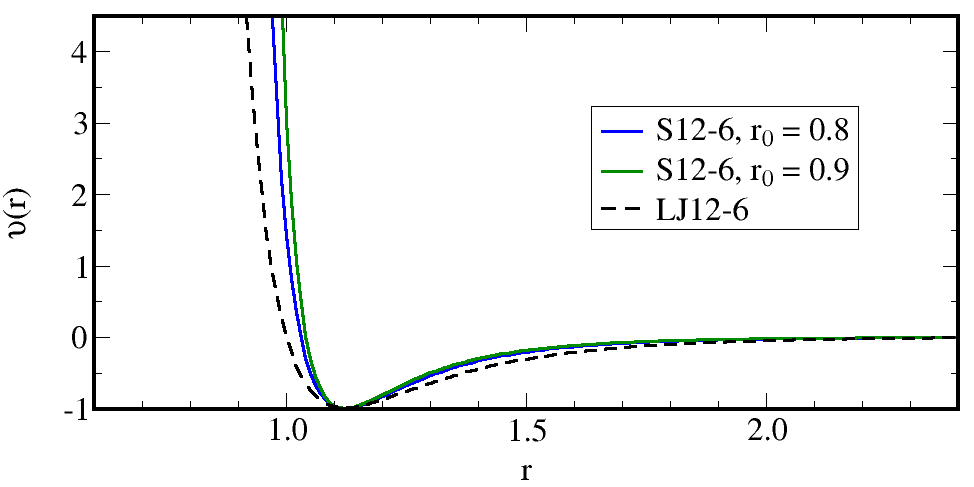}
  \caption{The S12-6 pair potential as described in Eq. (\ref{eq:alog_potential}) plotted for two values of $r_0$. The minimum was kept at $r_m = 2^{1/6}$ to make the two potentials comparable to the standard LJ12-6 potential, which is shown by the dashed line.}\label{fig:potentials}
\end{figure}

The S12-6 potential aims to reproduce the experimental \emph{change} in the potential steepness with density, not the steepness of the interatomic interaction itself. Taking the LJ12-6 as the starting point and ``adding'' a divergence at $r_0$, the S12-6 potential steeper than the LJ12-6 potential, and more so at short distances (see \fig{fig:potentials}). It is known that in many cases the LJ12-6 potential is too steep~\cite{Abrahamson1963, Halgren1992, Mattsson2010, Veldhorst2012}, so the same will be true for the S12-6. The steepness of the S12-6 potential around the minimum can however easily be tuned using the exponents $p$ and $q$.

\section{Simulation procedure\label{sec:method}}
The S12-6 pair potential is plotted in Fig.~\ref{fig:potentials}. To obtain densities comparable to the LJ liquids, the position of the potential minimum was chosen to be the same ($\varepsilon=1$ and $r_m=2^{1/6}$). This means that we use the LJ parameter $\sigma=2^{-1/6}r_m$ as the unit of length. Two values of the divergence diameter were simulated ($r_0=0.8\sigma$ and $r_0=0.9\sigma$). The ratio $r_0/r_m$ can be considered a measure for the steepness of the repulsion; in these cases it is 0.713 and 0.802, respectively. The potential was cut and shifted at $2.5\sigma$.

A cubic box with periodic boundary conditions and 1000 particles was simulated in the $NVT$ ensemble with a Nos{\'e}-Hoover thermostat. The integration time step was 0.001 for most state points, but it was decreased at high temperatures and densities in order to prevent unphysically large particle displacements. At each state point an initial configuration was first randomized by simulating $4 \times 10^6$ time steps at four times the desired temperature, followed by an equilibration run and a production run of $1.6 \times 10^7$ steps at the desired temperature.

The simulations were performed using the RUMD code~\cite{RUMD}. This code is optimized for performing Molecular Dynamics simulations on graphics processing units and is designed to make the implementation of new pair potentials straightforward~\cite{Bailey2015}.

\section{Results and discussion\label{sec:results}}
A configuration $\vect{R}$ is expressed in reduced (dimensionless) units by scaling with the density as $\rho^{1/3}\vect{R}$. Two state points $(\rho_1, T_1)$ and $(\rho_2, T_2)$ are defined as being isomorphic if configurations $\vect{R}_1$ and $\vect{R}_2$ of those state points with same reduced coordinates,
\begin{equation}\label{eq:scaled_conf}
  \rho_1^{1/3}\vect{R}_1 = \rho_2^{1/3}\vect{R}_2\,,
\end{equation}
also have proportional Boltzmann statistical weights~\cite{paper4}:
\begin{equation}\label{eq:isomorphDef}
          \exp\left(-\dfrac{U(\vect{R}_1)}{k_B T_1}\right) = 
  C_{1,2} \exp\left(-\dfrac{U(\vect{R}_2)}{k_B T_2}\right)\,.
\end{equation}
In practice, this proportionality should hold to a good approximation for most physically relevant configurations of the two state points with the same constant $C_{1,2}$ (which only depends on the pair of state points). Recently, a more general definition of the isomorph theory has been discovered~\cite{Schroder2014}, but we use here the ``older'' definition (\eq{eq:isomorphDef}), which arises from the new one via a first-order Taylor expansion and is more convenient for generating isomorphic state points in simulations.

The isomorph definition can be used to obtain a set of isomorphic state points from a simulation at an initial state point, by rewriting \eq{eq:isomorphDef} as
\begin{equation}\label{eq:dic}
  U(\vect{R}_2) = \dfrac{T_2}{T_1}U(\vect{R}_1) + k_BT_2\ln(C_{1,2})\,.
\end{equation}
Carrying out a standard equilibrium $NVT$ simulation at some initial state point 1, one calculates for each configuration first $U(\vect{R}_1)$ and subsequently $U(\vect{R}_2)$ by scaling to a new density $\rho_2$ using \eq{eq:scaled_conf}. According to \eq{eq:dic}, the energies of the scaled configurations should be linearly proportional to the energies of the initial configurations with proportionality constant $T_2/T_1$. In this way the temperature $T_2$ at which the state point at density $\rho_2$ is isomorphic to state point 1 is found from the slope in an  $U(\vect{R}_1)$, $U(\vect{R}_2)$ plot.

\begin{figure}
  \centering
  \includegraphics[width=\figwidth]{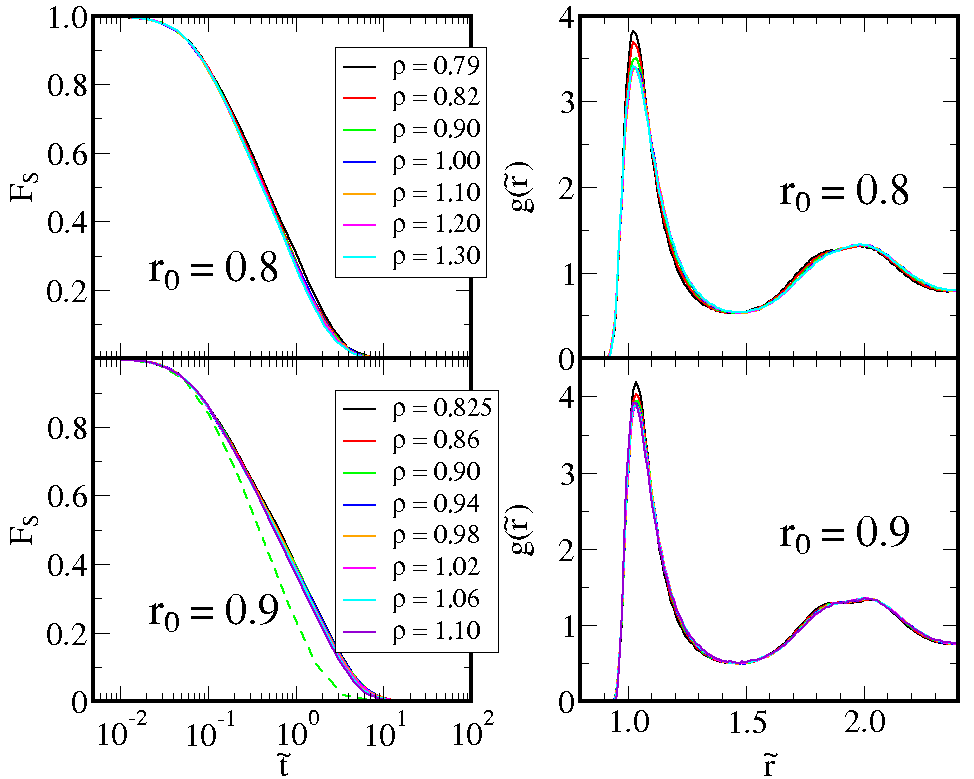}
  \caption{The incoherent intermediate scattering function (left) and the radial distribution function (right) in reduced units, for the potentials with hard-core diameter $r_0=0.8$ (top) and $r_0=0.9$ (bottom). For both isomorphs, the dynamics and structure are invariant to a high degree. All intermediate scattering functions have been calculated with the same reduced wave vector ($\tilde{q}=7.11$). The dashed green line shows the effect of an isochoris ($\rho=0.9$) temperature change for comparison.}\label{fig:alog_im_Fs_rdf}
\end{figure}

The intermediate scattering function $F_S(q,t)$ and the radial distribution function $g(r)$ along two isomorphs are plotted in \fig{fig:alog_im_Fs_rdf} in reduced units, defined as $\tilde{t} \equiv t\rho^{1/3}\sqrt{k_BT}$ and $\tilde{r}\equiv r\rho^{1/3}$. For $\rho=0.9$ and $r_0=0.9$ the effect on the dynamics of increasing the temperature by a factor of 2 while keeping density constant is also shown (dashed green line). The isochoric temperature change has a significant effect on the dynamics, whereas there is no significant change in $F_S$ along the isomorph, where temperature changes by a factor of 22. These data indicate that we have indeed obtained two sets of isomorphic state points, as both the dynamics and the structure are invariant in reduced units to a good approximation along the isomorph. There is a small change in the first peak of $g(r)$, though, which is expected, as at high density the particles are close and feel a steeper part of the potential. This leads to a steeper and therefore higher peak in $g(r)$~\cite{Veldhorst2015}.

\begin{figure}[t]
  \centering
  \includegraphics[width=\figwidth]{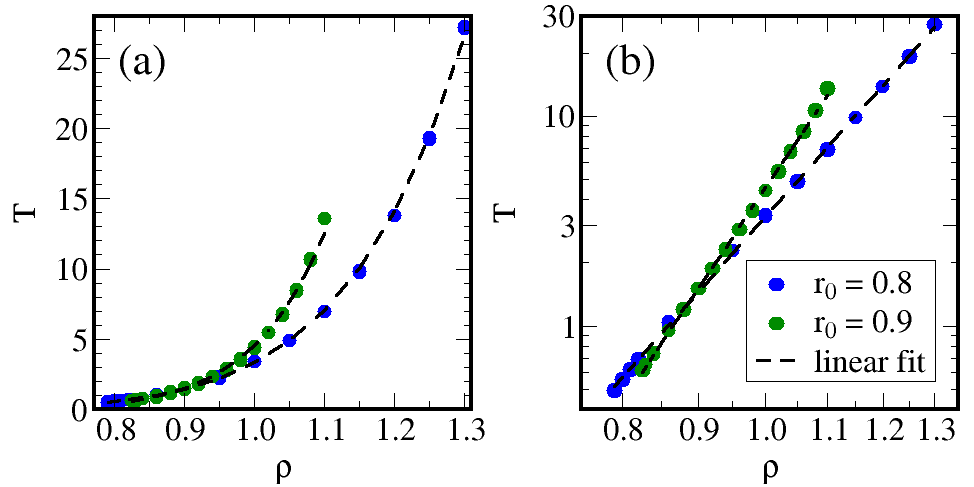}
  \caption{The shapes of the isomorphs in the $\rho,T$ phase diagram on a linear scale (a) and a log--log scale (b). The shapes of the ismorphs are dependent on the hard-core diameter $r_0$. On the log--log scale, the isomorphs appear to be fitted well by a straight lines (dashed lines).}\label{fig:alog_im_RhoT}
\end{figure}

The shape of the isomorphs in the $\rho,T$ plane is shown in \fig{fig:alog_im_RhoT} in a linear (a) and a double logarithmic scale (b). There is a clear effect of the particle diameter $r_0$. For $\gamma$ to be constant along the isomorph ($h(\rho) \propto \rho^\gamma$), the isomorph should be a  straight line in the log-log plot. There is barely any deviation from linear fits (dashed lines) in \fig{fig:alog_im_RhoT}(a), indicating that a power law is a good approximation. From the fits we find approximate ``constant $\gamma$'' values of 7.92 and 10.55 respectively for $r_0=0.8$ and $r_0=0.9$.

\begin{figure}[b]
  \centering
  \includegraphics[width=\figwidth]{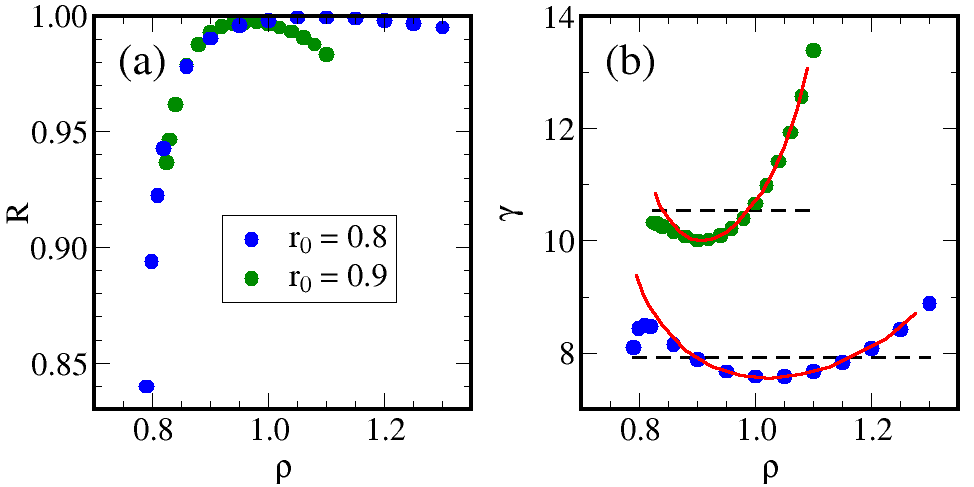}
  \caption{The correlation coefficient $R$ (a) and $\gamma$ (b) as a function of density, calculated from the fluctuations (\eq{eq:R} and \eq{eq:gamma-fluctuations}, green and blue dots). The red lines are finite difference estimates (see \eq{eq:gamma-hrho}) of the data in \fig{fig:alog_im_RhoT}. For both hard core diameters $r_0$ the $U,W$ correlations are strong, and $\gamma$ increases at high densities.}\label{fig:alog_im_Rgamma}
\end{figure}

As mentioned, liquids that obey the isomorph theory have strong correlations in the instantaneous values of the potential energy $U$ and the virial $W$. We investigate these correlations in \fig{fig:alog_im_Rgamma} plotting the correlation coefficient $R$ and the ``slope'' $\gamma$ from a linear regression of $U,W$ data (\eq{eq:gamma-fluctuations}). The liquids have strong correlations at all state points; only close to the gas--liquid coexistence region, at which the pressure becomes negative, is there a significant decrease in $R$. This is commonly found in many liquids when the pressure approaches zero~\cite{paper1}.

Although the isomorphs are fitted well by a straight line in \fig{fig:alog_im_RhoT}, the change in $\gamma$ is considerable when calculated from the logarithmic slope of $\rho,T$ data (red line). For both particle sizes we see an initial decrease in $\gamma$ with increasing density, similar to what is seen for liquids consisting of point particles like the LJ liquid~\cite{paper1}. However, unlike such standard systems, $\gamma$ increases again at higher densities. For $r_0=0.8$ the overall variation of the density-scaling exponent is small: $7.5 < \gamma < 8.5$, which is less than for the LJ potential considering the large range of densities. The potential is thus more in agreement with experimental density scaling data, although this seems to be caused by the cancellation of a low-density decrease and high-density increase in $\gamma$. The density increase is stronger when the hard-core radius is closer to the potential minimum; for $r_0=0.9$, $\gamma$ is clearly not constant, and there is a significant increase at higher densities.

Our results show the importance of the potential shape when large density changes are involved, especially when molecular liquids are simulated using coarse-grained models, as in this case it is common to use LJ potentials that diverge at zero distance~\cite{Shinoda2007, Riniker2012, Marrink2013}. We note that so far only one other pair potential gives an increasing $\gamma(\rho)$, which is the Girifalco potential~\cite{Girifalco1992, Bailey2013}. This potential was developed to model C$_{60}$ (Buckminsterfullerene), and the size of the the C$_{60}$ molecule led to a functional form that also diverges at a nonzero interparticle distance.

To conclude, our results for the S12-6 pair potential shed light on the discrepancy between experiments and simulations concerning the behavior of the density-scaling exponent $\gamma(\rho)$. We suggest that the decreasing $\gamma(\rho)$ seen in most simulations is a result of the potential used (often a LJ type potential). On the other hand, the constant value of $\gamma$ seen in experiments with molecular liquids seems to be an effect of the finite size of the molecules involved which is mimicked by the new S12-6 pair potential defined in \eq{eq:alog_potential}. Our results indicate that the size of molecules should be considered when choosing a pair potential to model a liquid over a large range of densities. Moreover, we predict that for large, bulky molecular liquids, the density-scaling exponent should increase at high densities, as was found in Ref.~\cite{Bohling2012}.

\acknowledgments{This work was sponsored by the Danish National Research Foundation via Grant No. DNRF61.}

\bibliography{log}
\end{document}